\documentclass[twocolumn,superscriptaddress,amsmath,aps]{revtex4}

\usepackage{amssymb,mathrsfs,bm,color,times}
\usepackage{graphicx,color}
\usepackage{bm}
\usepackage[hypertex]{hyperref}
\usepackage{amsmath}

\newcommand{\be}{\begin{equation}}
\newcommand{\ee}{\end{equation}}
\newcommand{\bea}{\begin{eqnarray}}
\newcommand{\eea}{\end{eqnarray}}
\newcommand{\bsube}{\begin{subequations}}
\newcommand{\esube}{\end{subequations}}

\newcommand{\Eq}[1]{Eq.\,(\ref{#1})}
\newcommand{\Eqs}[1]{Eqs.\,(\ref{#1})}

\newcommand{\la}{\langle}
\newcommand{\ra}{\rangle}

\newcommand{\beq}{\begin{equation}}
\newcommand{\eeq}{\end{equation}}
\newcommand{\beqn}{\begin{eqnarray}}
\newcommand{\eeqn}{\end{eqnarray}}

\newcommand{\nl}{\nonumber \\}

\newcommand{\bsub}{\begin{subequations}}
\newcommand{\esub}{\end{subequations}}

\begin{document}

\title{Postselected amplification applied to atomic magnetometers}

\author{Yazhi Niu}
\affiliation{Center for Joint Quantum Studies and Department of Physics,
School of Science, \\ Tianjin University, Tianjin 300072, China}

\author{Jialin Li}
\affiliation{Center for Joint Quantum Studies and Department of Physics,
School of Science, \\ Tianjin University, Tianjin 300072, China}

\author{Lupei Qin }
\email{qinlupei@tju.edu.cn}
\affiliation{Center for Joint Quantum Studies and Department of Physics,
School of Science, \\ Tianjin University, Tianjin 300072, China}

\author{Xin-Qi Li}
\email{xinqi.li@imu.edu.cn}
\affiliation{Center for Quantum Physics and Technologies,
School of Physical Science and Technology,
Inner Mongolia University, Hohhot 010021, China}
\affiliation{Center for Joint Quantum Studies and Department of Physics,
School of Science, \\ Tianjin University, Tianjin 300072, China}

\date{\today}

\begin{abstract}
{\flushleft We propose to embed }
the atomic magnetometer (AM) into an optical Mach-Zehnder interferometer (MZI).
We analyze the effect of amplification of the Faraday rotation (FR) angle
of the probe laser light, by properly postselecting
the path-information state of the laser photons when passing through the MZI.
In the presence of saturation of photo-detectors
and existence of polarization cross talk
in the polarizing-beam-splitter performance,
the amplified FR angle in the postselected photons
makes the scheme be able to
outperform the conventional measurement (without postselection),
being thus further enhancing
the sensitivity of the nowadays state-of-the-art optical AM.
\end{abstract}


\maketitle

{\flushleft\it Introduction}.---
Detection of weak magnetic fields plays important role in diverse fields
of fundamental science \cite{PXH21,PXH23,Rom22}
and practical technologies such as
in geology \cite{Ger20}, aeromagnetic investigation \cite{Fot20} ,
medical and military studies \cite{Bow22,Kaz20}, etc.
Accordingly, various techniques and instruments have been developed,
including such as fluxgate sensors \cite{Rip92,Rip22},
superconducting quantum interference devices (SQUIDs) \cite{Fag06},
Hall effect magnetic sensors \cite{Kar20},
and atomic magnetometers (AMs) \cite{Bud07,PXH21-b}.
Among these, SQUID may be the most popular and advanced one.
However, it works under extremely low temperatures,
requiring thus cryogenic maintenance
which causes, inevitably, expensive apparatus and operational costs.
In contrast, the full optical AM \cite{Bud07},
has the advantage of operating at or above room temperature
and getting smaller in volume with the potential for miniaturization.

Actually, the AMs have been developed for longer than half a century
\cite{Kas50,Deh57,Bloom57,Bloom62}.
In particular, along with the advent of tunable diode lasers
and progress of optical pumping techniques
and producing dense atomic vapours
with long ground-state spin relaxation times (in some cases $\sim$ 1 sec),
the atomic magnetometers have achieved sensitivities
comparable to or even surpassing
that of most SQUID-based magnetometers,
having thus become a leading magnetometry of
ultrasensitive magnetic field measurements \cite{Bud17}.
By constantly improving,
the most advanced atomic magnetometers
have stepped at the precision level of subfemtotesla \cite{Bud00,Rom02,Rom10}.

In recent experiments \cite{PXH21,PXH23,Rom22},
the performance of the $^{87}$Rb AM was improved
by mixing $^{129}$Xe gas into the $^{87}$Rb gas in the same vapor cell,
with the $^{129}$Xe nuclear spins
playing a role of a pre-amplifier of weak magnetic fields.
The amplified {\it effective magnetic field}
acting onto the electron spins of the $^{87}$Rb atoms
can be two orders of magnitude larger than the original field.
This new technique has also been discussed to probe signals from such as
Goldstone bosons of new high-energy symmetries,
CP-violating long-range forces, and axionic dark matter, etc.

\begin{figure}
  \centering
  \includegraphics[scale=1.2]{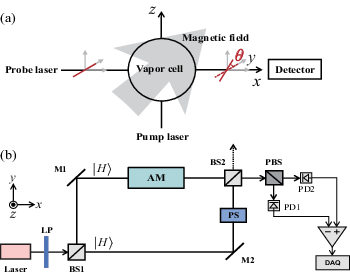}
  \caption{
(a)
Schematics of principle of the atomic magnetometer (AM).
Atoms (e.g. $^{87}$Rb) in the vapor cell are polarized
under the illumination of a pump laser light (circularly polarized),
and a linearly polarized probe laser light
is propagated along the $x$-direction.
The probe light suffers an optical Faraday rotation (FR)
when passing through the vapor cell.
Measuring the FR angle allows to infer an unknown magnetic field.
(b)
Proposal to insert the AM into one of the arms of
an optical Mach-Zehnder interferometer (MZI).
Proper postselection for the ``path-state" of the probe light
can realize the effect of postselected amplification (PSA) for the FR angle.
Abbreviations of device elements in the setup:
LP (linear polarizer),
BS (beam splitter),
PBS (polarizing beam splitter),
PS (phase shifter),
PD (photodiode),
M (mirror),
and DAQ(data acquisition).      }
\label{fig1}
\end{figure}

In this work, we propose a postselected metrological scheme
to further enhance the probe sensitivity of
the $^{87}$Rb AM (the principle is also applicable for other AMs).
The idea of postselection was recently reanalyzed
in a broader context \cite{Llo20,Shu22,Yang23},
as a generalization of the weak-value amplification (WVA) \cite{AAV1,AAV2}.
The technique of WVA has been successfully demonstrated in precision measurements
\cite{Kwi08,How09a,How10a,How10b,How13,Sto12,Lun17,ZLJ20},
showing profound advantages in the presence of detector's saturation,
technical noises, and some environmental disturbances.
Our proposal is suggesting to insert the AM into an optical
Mach-Zehnder interferometer (MZI) (see Fig.\ 1),
which renders the photons of the probe laser light
having a path-information (``which path") degree of freedom
in addition to the optical polarization degree of freedom.
Then, by properly postselecting the path information
from the path-and-polarization entangled state
(owing to path-dependent Faraday rotation),
we will show the effect of postselected amplification (PSA)
of the Faraday rotation (FR) angle.
This technique is anticipated to
benefit the nowadays AM in probing tiny magnetic fields.

\vspace{0.2cm}
{\flushleft\it Scheme of Postselected Amplification}.---
The basic principle of an AM (e.g., the $^{87}$Rb AM) consists of
two operations, as schematically shown in Fig.\ 1(a).
First, the $^{87}$Rb atoms in the vapor cell are polarized
under the illumination of a pump laser light (circularly polarized).
The pump laser couples the outmost valence electron of the $^{87}$Rb atom
to resonant transition between $5^2S_{1/2}$
and $5^2P_{1/2}$ (i.e., the D1 line at 794.98 nm).
After experiencing a few dynamic processes,
most atoms are transferred to
the $m_J=+1/2$ (actually $m_S=+1/2$)
ground state along the $\hat{z}$-axis,
with the $^{87}$Rb atoms in the vapor cell
having a polarization vector $\textbf{P}=P_{z,0} \textbf{e}_z$.
Second, let the polarization vector $\textbf{P}$
precess around a magnetic field $\textbf{B}$,
and propagate a linearly polarized probe laser light
along the $\hat{x}$ direction,
which is detuned from the D2 transition line at 780.2 nm
between $5^2S_{1/2}$ and $5^2P_{3/2}$.
The probe light will suffer an optical FR
after passing through the atom vapor cell.
The FR angle is proportional to the $x$-component of $\textbf{P}$.
Then, measuring the FR angle allows us to infer
the unknown magnetic field, which is contained in $\textbf{B}$.

In the above description, for simplicity, we ignored the effect of nuclear spins.
If including the nuclear spin $I=3/2$ of the $^{87}$Rb atom
and accounting for its coupling
with the electron's angular momentum, say, ${\bf J=L+S}$ and ${\bf F=J+I}$,
the hyperfine states with $F=1$ and $F=2$ have Zeeman splittings
between the degenerate hyperfine sublevels
$m_F=0, \pm 1$ and $m_F=0, \pm 1, \pm2$.
Under the illumination of a pump laser light
(circularly polarized, along the $z$ direction),
after a few dynamic processes (collision and relaxation processes),
the $m_F=+2$ ground state will become primarily populated,
since this state is transparent to the pump laser beam,
i.e., electron on this state cannot absorb photon
to transit to any of the excited sublevel states.
Notice that, in the $m_F=+2$ ground state,
the electron spin is $m_S=+1/2$.
Thus, the electron spin polarization vector $\textbf{P}=P_{z,0} \textbf{e}_z$
will be resulted in (for more details, see Appendix A).

Now, consider to insert the AM into one of the arms of an optical MZI,
as schematically shown in Fig.\ 1(b).
The polarized probe light is an ensemble of photons with the same
polarization and the same frequency.
We thus adopt the quantum mechanical description of single photon for the probe light.
For the setup under consideration, the transverse spatial wavefunction
of the light beam is irrelevant to the problem,
therefore each single photon has two degrees of freedom:
one is the polarization, which can be described
using the basis states $|H\ra$ and $|V\ra$;
another is its path information when passing through the MZI,
using $|1\ra$ and $|2\ra$ to denote it.
We assume that the probe light is prepared initially
with the polarization of $|H\ra$.
After entering the MZI through the first beam splitter (BS1),
the path state of the photon is $|i\ra=(|1\ra + |2\ra)/\sqrt{2}$.
Only propagating along path ``1", the photon's polarization $|H\ra$ will suffer
a Faraday rotation of angle $\theta$,
owing to magnetic polarization of the electron spins of the atomic medium in the vapor cell.
In contrast, if the photon propagates along path ``2",
its polarization remains unchanged.
This path-dependent Faraday rotation can be elegantly described as
\bea\label{entang-state}
|\Psi(\theta)\ra = e^{-i\theta\hat{A}\hat{\sigma}_x} |i\ra|H\ra \,.
\eea
Here we introduced $\hat{A}=|1\ra \la 1|$,
and $\hat{\sigma}_x$ is the Pauli operator
defined in the Hilbert space expanded by $\{|H\ra, |V\ra\}$.

To realize the effect of PSA, let us assume to measure only the light
outgoing from the exit port along $x$-direction
through the second beam-splitter of the MZI (BS2, see Fig.\ 1(b)),
and to extract the FR angle
by a difference-signal data analyzer
(i.e., the DAQ shown in Fig.\ 1(b)).
This {\it one-exit-port} measurement (discarding the light in the other exit port)
corresponds to a postselection of the path state.
More specifically, the state of postselection is
$|f\ra = (|1\ra + e^{i\beta}|2\ra)/\sqrt{2}$.
Here the phase factor $e^{i\beta}$ is introduced
by a phase shifter which is inserted in path ``2", as shown in Fig.\ 1(b).
Via modulating $\beta$, one can realize
an {\it effect of PSA} for the FR angle.
Below we outline the key treatment for achieving this desired effect.

After postselection with $|f\ra$,
the polarization state of the postselected photon becomes
\bea\label{ps-state}
|\Phi_f(\theta)\ra
= \frac{1}{2\sqrt{p_f}}   \left[(e^{-i\beta}+\cos\theta)|H\ra
-i\sin\theta |V\ra \right] \,.
\eea
Mathematically, the postselection is described by
$|\widetilde{\Phi}_f(\theta)\ra = \la f|\Psi(\theta) \ra$.
$|\Phi_f(\theta)\ra$ is the normalized version of
$|\widetilde{\Phi}_f(\theta)\ra$,
where the normalization factor is associated with
the postselection probability $p_f$, which is given by
$p_f=\la \widetilde{\Phi}_f(\theta)|\widetilde{\Phi}_f(\theta)\ra
= (1+ \cos\theta \cos\beta)/2$.

It would be instructive to get a preliminary insight
for the effect of amplification through postselection,
by computing first the quantum Fisher information (QFI) about $\theta$,
contained in $|\Psi(\theta)\ra$ and $|\Phi_f(\theta)\ra$, respectively.
For $|\Psi(\theta)\ra$, simple calculation gives the QFI
about $\theta$ \cite{Cra16,Rao92}, as ${\cal I}(\theta)=2$.
However, remarkably, for $|\Phi_f(\theta)\ra$,
the QFI is obtained as
\bea
{\cal I}^{\rm ps}(\theta) = \frac{4p_f-\sin^2\theta}{4p^2_f}  \,.
\eea
One can check that, ${\cal I}^{\rm ps}(\theta)$
can be anomalously larger than $2$.
This means that each postselected photon carries more QFI
than the photon before postselection, in the entangled state $|\Psi(\theta)\ra$.
This result reflects the key feature of
the recently advocated postselection filtering technique,
which has been termed as {\it postselected metrology} \cite{Llo20}.
The origin of the anomalous QFI has been profoundly connected with
the unique quantum nature
of {\it negativeness} of the Kirkwood-Dirac quasiprobability,
and it was pointed out that the postselected metrology can
outperform the optimal postselection-free experiment \cite{Llo20}.
Indeed, in a subsequent experiment \cite{Shu22},
the metrological advantage of this technique was demonstrated,
showing that it can amplify the information
by two orders of magnitude, in per postselected photon.

\begin{figure}
  \centering
  \includegraphics[scale=0.5]{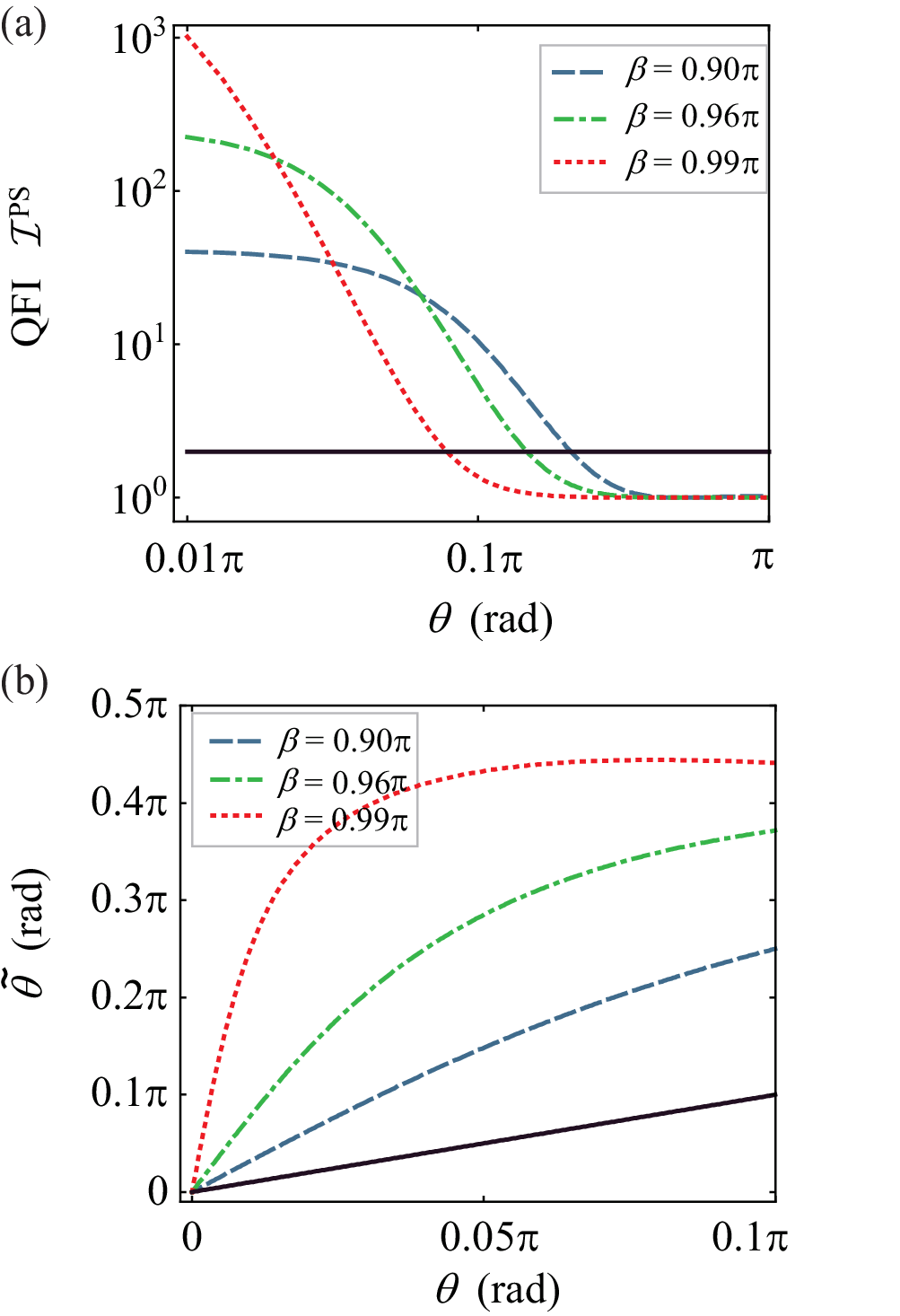}
  \caption{
(a)
The QFI ${\cal I}^{\rm ps}(\theta)$,
carried by per postelected photon in state $|\Phi_f(\theta)\ra$,
under different choices of the postselection parameter $\beta$,
compared with the QFI encoded
in the entangled state $|\Psi(\theta)\ra$
given by \Eq{entang-state} (plotted here by the black line).
(b)
The PSA-FR angle $\widetilde{\theta}$
{\it versus} the true FR angle $\theta$, under different choices of $\beta$.
The slope of each curve in the small $\theta$ regime
characterizes the amplification factor, say, $\widetilde{\theta}/\theta$.
Compared with the black line (result without postselection),
the PSA effect is evident.        }
\label{fig2}
\end{figure}

In Fig.\ 2(a), for our setup, we display the QFI ${\cal I}^{\rm ps}(\theta)$
carried by per postelected photon under different choice of
the postselection parameter (say, the phase shift parameter $\beta$).
We see that, drastically, ${\cal I}^{\rm ps}(\theta)$ can exceed
the QFI encoded in the entangled state $|\Psi(\theta)\ra$,
i.e., ${\cal I}(\theta)=2$.
${\cal I}^{\rm ps}(\theta)$ can exceed also the QFI
associated with the conventional measurement (CM),
shown in Fig.\ 1(a),
which encodes the FR angle $\theta$
in the polarization state of a probing photon
through the transformation
$|\Phi^{\rm cm}(\theta)\ra = e^{-i\theta \hat{\sigma}_x} |H\ra$,
which carries the QFI ${\cal I}^{\rm cm}(\theta)=4$.
Notice that, ${\cal I}(\theta)$ is only a half of ${\cal I}^{\rm cm}(\theta)$.
The reason is that, when a photon transmits through the MZI,
it is affected by the parameter $\theta$ with only a half probability.

\vspace{0.2cm}
{\flushleft\it Measurement and Signal Analysis}.---
Below we continue our analysis for practical measurement
of the amplified signal of the FR angle.
In conventional measurement,
the FR angle $\theta$ is extracted from the probability $P_V=\sin^2\theta$,
of the $|V\ra$ component in the state $|\Phi^{\rm cm}(\theta)\ra$,
which is generated by the FR transformation.
In the postselected measurement, based on \Eq{ps-state},
the probability of the $|V\ra$ component
in the postselected state $|\Phi_f(\theta)\ra$ is given by
\bea\label{PSA-FR}
\widetilde{P}_V(\theta)
= \sin^2\theta /4p_f \equiv \sin^2\widetilde{\theta}  \,.
\eea
Here, by analogy with conventional measurement,
we define an effective angle $\widetilde{\theta}$,
which is termed as PSA-FR angle hereafter in this work.
This result indicates that, with the decrease of $p_f$,
$\widetilde{\theta}$ can be much larger than the true FR angle $\theta$,
realizing thus a remarkable amplification effect.
In Fig.\ 2(b), we illustrate how the FR angle $\theta$ is amplified to $\widetilde{\theta}$,
by modulating the postselection parameter $\beta$.
For smaller $\theta$, the amplification effect can be more prominent,
for instance, it can be amplified by several orders of magnitude.

In practice, $\widetilde{P}_V$ can be obtained
by measurement as follows.
That is, the postselected photons are guided to a PBS (polarizing beam splitter),
as shown in Fig.\ 1(b),
which is set to allow the $|H\ra$-component to transmit through it,
and the $|V\ra$-component to be reflected.
Then, measure the photon flux intensities
of the $|H\ra$ and $|V\ra$ polarizations, by two photo-detectors,
which yield photo-currents $I_H=I_0 \widetilde{P}_H$
and $I_V=I_0 \widetilde{P}_V$, respectively.
The difference-signal data analyzer, i.e., the DAQ shown in Fig.\ 1(b),
gives the ratio
\bea\label{DAC}
R=(I_V-I_H)/(I_V+I_H)=2\widetilde{P}_V -1  \,,
\eea
from which we straightforwardly obtain the key quantity $\widetilde{P}_V$.    \\

Below we further specify our consideration to connect with
the real $^{87}$Rb atomic magnetometer,
which was improved in a recent experiment \cite{PXH21},
by mixing $^{129}$Xe gas into the $^{87}$Rb gas in the same vapor cell,
and exploiting the $^{129}$Xe nuclear spins
playing role of a pre-amplifier to measure weak magnetic field.
It was demonstrated in Ref.\ \cite{PXH21} that
the amplified {\it effective magnetic field}
acting onto the electron spins of the $^{87}$Rb atoms
can be two orders of magnitude larger than the original field.
Thus, roughly speaking, the FR of polarization of the probe light
is amplified by two orders of magnitude.
This amplification technique has been discussed to be useful in probing
extremely weak magnetic effects, including such as searching dark matters.

Below, using the real parameters of the $^{87}$Rb AM,
let us carry out an estimation
for the time dependent FR angle $\theta(t)$.
We assume that
the electron spins of the $^{87}$Rb atoms
are polarized in the $z$-direction,
see Fig.\ 1(a), by the pump laser light,
and that a bias magnetic field (relatively large)
is applied along the $z$-direction.
Then, in the absence of any other magnetic fields,
the polarization vector $\textbf{P}$
does not precess around the bias magnetic field.
However,
if a weak magnetic field (to be estimated) is present,
e.g., in the $y$-direction,
the polarization vector $\textbf{P}$
will precess around the total magnetic field.
As a result, the component $P_x(t)$ of $\textbf{P}(t)$ in the $x$-direction
will cause an optical rotation of the linearly polarized probe laser light
after transmitting through the vapor cell (along the $x$-direction),
with the FR angle proportional to $P_x(t)$ as \cite{PXH21}
\begin{equation}\label{FR-theta}
  \theta(t)=\frac{1}{4}l r_{e} c f n D(\nu) P_{x}(t)  \,.
\end{equation}
In this result, $l$ is the optical path length of the probe light
transmitting through the vapor cell;
$r_{e}=e^2(4\pi\varepsilon_0 m_e c^2)^{-1}=2.8$ fm
is the classical radius of electron;
$c$ is the speed of light;
and $f$ is the oscillator strength of electron
(about 1/3 for the D1 transition and 2/3 for the D2 transition).
The Lorentzian lineshape function,
$D(\nu)=(\nu-\nu_{0})/{[(\nu-\nu_{0})^{2}+(\Delta\nu/2)^{2}]}$,
characterizes the effect of detuning
between $\nu$ (frequency of the probe light)
and $\nu_{0}$ (frequency of the D2 transition).
In the real experiment \cite{PXH21},
the probe light is blue-detuned by 110 GHz from the D2 transition.
$\Delta\nu$ is the full-width at half-maximum (FWHM),
owing to level broadening,
or the damping rate of oscillating dipole in classical treatment.
The off-resonance D2 transition,
or the respective classical electric polarization of the atoms,
causes a change of the refractive index $n(\nu)$
of the atomic medium to the light interacting with it.
If $P_x\neq 0$, which indicates
the electron spin probabilities $p_{+1/2}\neq p_{-1/2}$
(the spin component is projected along the $x$ direction),
then the left ($L$) and right ($R$) circularly polarized lights
will cause different electric polarizations of the atoms
(owing to the selection rule of angular momentum conservation),
resulting thus in different refractive indices, $n_{+}(\nu)\neq n_{-}(\nu)$.
From $\theta=\frac{\pi\nu l}{c}[n_+(\nu)-n_-(\nu)]$,
one obtains the result of \Eq{FR-theta}.

To carry out specific result of the FR angle,
we need to obtain $P_x(t)$, from the following Bloch equation \cite{Hap98,Sel08,The12}
\begin{equation}\label{Bloch}
  \frac{d}{dt}\textbf{P}= \frac{1}{q} [ \gamma_{e}\textbf{B}\times \textbf{P}
  +R_{\rm op}(\textbf{e}_z -\textbf{P})-R_{\rm rel}\textbf{P} ]    \,.
\end{equation}
In this equation, $\textbf{P}$ is the Bloch vector of the electron spin;
$\gamma_{e}=g_{s}\mu_{B}$,
with $g_{s}\approx 2$ the electron's Land$\acute{e}$ $g$ factor
and $\mu_{B}$ the Bohr magneton;
$R_{\rm op}$ is the optical pumping rate
and $R_{\rm rel}$ is the spin relaxation rate.
Moreover, $\textbf{e}_z$ is the unit vector in the direction of the pumping light,
and $\textbf{B}$ is the total magnetic field.
The above Bloch equation was derived from
a joint description for the electron and nuclear spins,
and tracing out the degrees of freedom of the nuclear spins
after the hyperfine interaction \cite{Hap98}.
The reduction factor (also called `nuclear slowing-down factor' in literature)
is defined as $q=\la F_z\ra / \la S_z\ra$,
with $\la F_z\ra$ the quantum average of the $z$-component of
the total angular momentum ${\bf F}$
(${\bf F=J+I}$, the sum of the electron's total angular momentum ${\bf J}$
and the nuclear spin ${\bf I}$),
and $\la S_z\ra$ the quantum average of the $z$-component of
the electron spin ${\bf S}$.
For the $^{87}$Rb atom, the nuclear spin $I=3/2$,
it was found \cite{Hap98,Sel08}
that the $q$ factor is $q = 2(3+P^2)/(1+P^2)$,
with $P$ defined as $P\equiv P_z=2\la S_z\ra$.
Similarly, the polarization vector
${\bf P}$ is defined as ${\bf P}=2 \la{\bf S}\ra$.     

Following the experiment \cite{PXH21},
we assume a relatively large constant bias field $B_z(t)=B_z$
applied in the $z$-direction,
and a weak alternating field, $B_y(t)=B_y\cos(2\pi\nu t)$,
to be measured, in the $y$-direction.
Under the condition $R_{\rm op}\gg \gamma_e B_y$,
one can first obtain the approximate solution
$P_z (t)\simeq P_{z,0}=R_{\rm op}/(R_{\rm op}+R_{\rm rel})$,
then  obtain $P_x(t)={\cal M}  P_{z,0} B_y \cos(2\pi\nu t+\theta_y)$.
Both the amplitude modulation factor ${\cal M}$
and the phase delay factor $\theta_y$
are $B_z$ dependent, but have lengthy expressions (thus not shown here).

\begin{figure}
  \centering
  \includegraphics[scale=0.55]{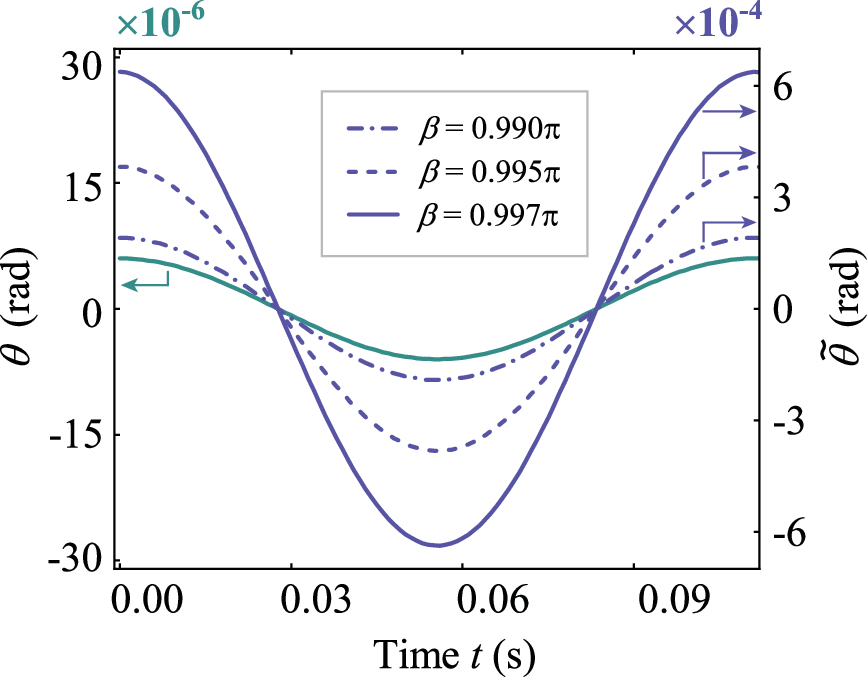}
  \caption{
Time dependent FR signals,
by comparing the true FR angle $\theta$
(dark-green curve, left coordinate)
with the PSA-FR angle $\widetilde{\theta}$
(purple curves, right coordinate).
The results are simulated using parameters referred to Ref.\ \cite{PXH21},
which are summarized in the main text of present work.      }
\label{fig3}
\end{figure}

In Fig.\ 3, we display the numerical result of PSA of the FR angle,
for a few choices of the postselection parameter $\beta$,
using parameters
referred to the experiment \cite{PXH21}.
Some parameters not specified below \Eq{FR-theta} are:
the optical path length $l=1$ cm,
the density of atom numbers $n=14~{\rm cm}^{-3}$,
and the FWHM $\Delta \nu=59$ Hz.
The values of the quantities in the Bloch equation are:
the gyromagnetic ratio
$\gamma_{\rm e}=2\pi\times 28 ~{\rm Hz/nT}$,
the pumping rate $R_{\rm op}= 2800 ~{\rm sec}^{-1}$
and the relaxation rate
$R_{\rm rel}= 1000 ~{\rm sec}^{-1}$,
the magnetci fields $B_z=759 ~{\rm nT}$ and
$B_y= 30\times \cos(2\pi\times 8.96 t)~{\rm pT}$.
Using these parameters,
our simulation gives the result shown in Fig.\ 3.
The effect of amplification is prominent.
For example, for the postselection using $\beta=0.997\pi$,
the amplification factor $\eta=\widetilde{\theta}/\theta$ is larger than 100.\\
\\

\vspace{0.1cm}
{\flushleft\it Discussion and Summary}.---
The precision of an AM is largely affected by shot noise
of the discrete photon numbers of the probe light beam.
Let us consider the quantity $R$, measured in real experiment, given by \Eq{DAC}.
We may denote the collected photon numbers in the two photo-detectors as
$N_1=p_f N \cos^2\widetilde{\theta}$ and $N_2=p_f N \sin^2\widetilde{\theta}$.
Then, we have $R=(N_1-N_2)/(N_1+N_2)$.
In general, the shot noise (uncertainty of the photon numbers)
is $\delta N_1=\sqrt{N_1}$ and $\delta N_2=\sqrt{N_2}$.
The error of $R$, originated from $\delta N_1$ and $\delta N_2$, is determined by
$\delta R=[(\frac{\partial R}{\partial N_1}\delta N_1)^2
+ (\frac{\partial R}{\partial N_2}\delta N_2)^2 ]^{1/2}$.
Simple algebra gives
$\delta R=\sin 2\widetilde{\theta}/\sqrt{p_f N}$.
Based on \Eqs{DAC} and (\ref{PSA-FR}),
we further obtain $\delta \widetilde{\theta} = 1/(2\sqrt{p_f N})$.

We see that the signal-to-noise ratio of the PSA scheme,
$\widetilde{\cal R}_{S/N}=\widetilde{\theta}/\delta\widetilde{\theta}$,
remains the same as the conventional measurement,
${\cal R}_{S/N}= \theta/\delta\theta$,
by noting that
$\widetilde{\theta}\simeq \theta/(2\sqrt{p_f})$
(in the case of small $\theta$ and $\widetilde{\theta}$)
and $\delta\theta \sim 1/\sqrt{N}$.
This implies that the small number of postselected photons
contain almost all the metrological information
of all the photons before postselection.
As a result, the PSA provides an approach
to ensure that the detector operates under the
saturation threshold even for a large number of input photons,
leading thus to remarkably outperforming the conventional measurement,
as analyzed in theory \cite{Lun17}
and demonstrated by experiment \cite{ZLJ20}.
For the case of AM, in the presence of saturation
of the two photo-detectors (PD1 and PD2 in Fig.\ 1),
one can expect that increasing the intensity of the probe light can make
the PSA scheme remarkably outperform the conventional measurement without postselection.

Moreover, the PSA metrology has been proposed and demonstrated
to amplify miniscule physical effects,
holding the potential for enhancing measurement sensitivity
and overcoming some technical imperfections.
In our case, for the FR-based optical AM,
one of the important technical issues
is the existence of {\it polarization cross talk}
in the PBS performance \cite{Van11,Wang13},
in the last stage of measuring the polarization component ratio caused by FR.
In ideal case (theoretical model), if there is no polarization cross talk,
the $|V\ra$ component of the light
will fully reflected (deflected) through the PBS and enter PD1,
while the $|H\ra$ component
will be fully transmit and enter PD2, as schematically shown in Fig.\ 1(b).
Then, after a procedure of calibration \cite{Wang13,Ben01,Win06,Yang19}
(see Appendix B, for some details),
the measured ratio $\widetilde{V}_{\rm m}$
just recovers the true ratio $V_0=P_V/P_H$
in the light before entering the PBS.
Precisely (reliably) measuring $V_0=P_V/P_H$
is one of the key ingredients in the FR-based full optical AM,
which is also the key task in the polarization lidar \cite{Win06,Yang19,Bau00,Li23}.
In real PBS system, inevitably, there exists polarization cross talk.
This imperfection will affect the ultimate limit of $V_0$ measurement (estimation),
and has received extensive studies
in the context of polarization lidar \cite{Win06,Yang19,Bau00,Li23}.
In Appendix B, we show that in the presence of polarization cross talk,
the $V_0$ measurement/estimation error
will become more and more serious with the decreasing of $V_0$.
Remarkably and desirably,
the key point of PSA applied to the FR-based optical AM
is amplifying the FR angle, from $\theta$ to $\widetilde{\theta}$,
as shown in Fig.\ 2(b), for instance, by larger than two orders of magnitude.
This implies similar orders of magnitude
amplification of $V_0$, in the small $V_0$ limit.
Therefore, the PSA strategy proposed in this work
is anticipated to be very useful to
enhance the ultimate limit of weak magnetic fields probing,
by means of the FR-based optical AM.           

To summarize, we analyzed in this work the effect of amplification of the FR angle
via properly postselecting the path state,
by considering to embed the optical AM into an MZI.
Physically speaking, this is because the postselected photon,
compared with the photon in the conventional measurement without postselection,
contains much more information about the FR angle,
while the FR angle is related with (proportional to)
the weak magnetic field to be estimated.
Therefore, the scheme proposed in this work provides
a postselected amplification metrology of probing tiny magnetic fields,
holding the advantage of further enhancing the sensitivity of the nowadays optical AM.
Especially, in the presence of saturation of photo-detectors
and existence of polarization cross talk in the PBS performance,
the proposed scheme is expected to remarkably
outperform the conventional measurement (without postselection).

\vspace{0.5cm}
{\flushleft\it Acknowledgements.}---
This work was supported by
the NNSF of China (Nos.\ 11675016, 11974011 \& 61905174).

\appendix
\section{The effect of nuclear spins}

In the main text, for simplicity, we ignored the effect of nuclear spins.
For $^{87}{\rm Rb}$ atom, the single electron in the valence shell
is in the $5 ^2S_{1/2}$ state as its ground state.
The pump laser light is tuned to the $D_1$ line,
making transition between $5 ^2S_{1/2}$ and the excited state $5 ^2P_{1/2}$.
If including the nuclear spin $I=3/2$ and accounting for its coupling
with the electron's angular momentum, say, ${\bf J=L+S}$ and ${\bf F=J+I}$,
the hyperfine states with $F=1$ and $F=2$ have Zeeman splittings
between the degenerate hyperfine sublevels
$m_F=0, \pm 1$ and $m_F=0, \pm 1, \pm2$, respectively, see Fig.\ 4

Consider optical pumping with $\sigma^+$ circularly polarize light, along the $z$-direction.
Then, all the photons have angular momentum $+1$ along this direction.
According to the rule of conservation of angular momentum,
the ground state sublevels with $m_F$ are depopulated
and the excited state sublevels with $m'_F=m_F+1$ are populated, as shown in Fig.\ 4.
Since the $^{87}$Rb vapor cell contains ${\rm N}_2$ buffer gas,
together with the Rb-Rb collisions,
collisional quenching of the electron from excited to ground states
and collisional mixing of sublevel states would be induced.
The relaxation processes are random.
Therefore, eventually, the $m_F=+2$ ground state becomes primarily populated,
since this state is transparent to the incident pump laser beam,
i.e., electron on this state cannot absorb photon
to transit to any of the excited sublevel states (see Fig.\ 4).
Generally speaking, the relative rates of relaxation and optical pumping
determine the steady-state polarization.
In practice, a polarization on the order of $1/2$ is ideal.
To achieve the $1/2$ polarization, the relaxation rate should equal the pumping rate.
This requires the power strength of the pump laser to be adjusted accordingly.

\begin{figure}
  \centering
  \includegraphics[scale=0.5]{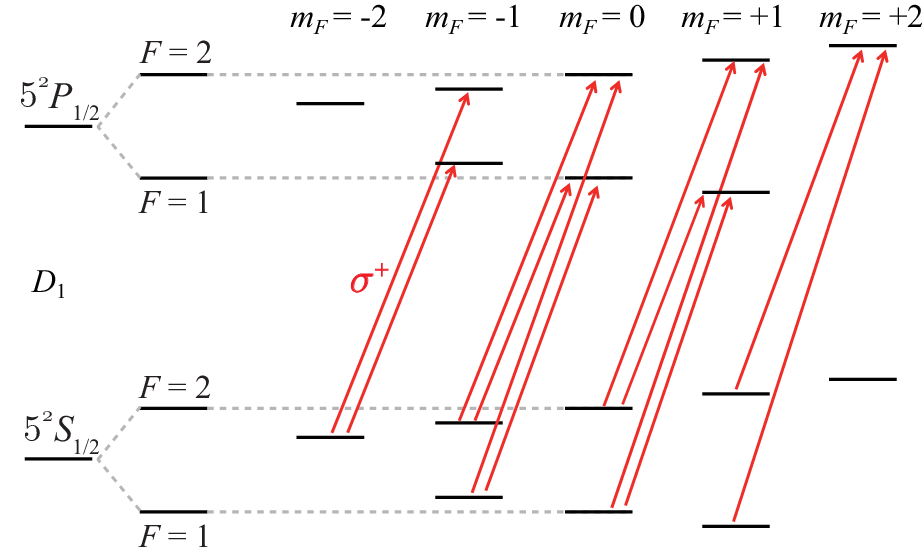}
  \caption{
Including the nuclear spin $I=3/2$ of the $^{87}{\rm Rb}$ atom
and accounting for its coupling with the electron's angular momentum,
the hyperfine states with $F=1$ and $F=2$
of the states $5 ^2S_{1/2}$ and $5 ^2P_{1/2}$,
have Zeeman splittings between the degenerate hyperfine sublevels
$m_F=0, \pm 1$ and $m_F=0, \pm 1, \pm2$.
The pump laser light (with $\sigma^+$ circular polarization)
is tuned to the $D_1$ line,
making transitions such that
the ground state sublevels with $m_F$ are depopulated
and the excited state sublevels with $m'_F=m_F+1$ are populated.
After a few collision and relaxation processes,
the $m_F=+2$ ground state becomes primarily populated,
since this state is transparent to the pump laser beam.
As a result, the electron's spin in the ground state $5 ^2S_{1/2}$ is partially polarized.   }
\label{fig2}
\end{figure}

In particular, if the spin relaxation rate (collision caused)
between the ground-state sublevels
is much larger than the optical pumping rate,
the occupations of the ground-state sublevels
satisfy the so-called ``spin-temperature" distribution \cite{The12}:
$\rho(m_F)=e^{\beta m_F}/Z$,
with $Z$ the partition function, $Z=\sum_{m_F} e^{\beta m_F}$,
and the inverse temperature
$\beta = \ln [(1+P_z)/(1-P_z)]$.
For instance, for $P_z=0.5$,
the relative occupation probabilities, $\tilde{\rho}(m_F)=e^{\beta m_F}$,
of the $m_F=-2, -1, 0, +1$ and $+2$ ground-state sublevles
are 0.11, 0.33, 1, 3 and 9, respectively.

Starting with a joint description for the electron and nuclear spins,
and tracing out the degrees of freedom of the nuclear spins
after the hyperfine interaction,
one can derive an elegant reduced Bloch equation as \cite{Hap98}
\begin{equation}\label{Bloch}
  \frac{d}{dt}\textbf{P}= \frac{1}{q} \left[ \gamma_{e}\textbf{B}\times \textbf{P}
  +R_{\rm op}(\textbf{e}_z -\textbf{P})-R_{\rm rel}\textbf{P} \right]  \,.  \nonumber
\end{equation}
In this result, the reduction factor is defined as $q=\la F_z\ra / \la S_z\ra$,
with $\la F_z\ra$ the quantum average of the $z$-component of
the total angular momentum ${\bf F}$,
and $\la S_z\ra$ the quantum average of the $z$-component of
the electron spin ${\bf S}$.
Notice that, in the ground state $5^2S_{1/2}$, the orbital angular momentum is zero.
For the $^{87}$Rb atom, the nuclear spin $I=3/2$,
it was found \cite{Hap98,Sel08} that the $q$ factor is $q = 2(3+P^2)/(1+P^2)$,
with $P$ defined as $P\equiv P_z=2\la S_z\ra$.
Similarly, the polarization vector
${\bf P}$ is defined as ${\bf P}=2 \la{\bf S}\ra$.
From the reduced Bloch equation,
we see that the $q$ factor does not affect the steady-state solution,
i.e., in the absence of magnetic field,
the steady-state solution reads as
$P_z = R_{\rm op}/(R_{\rm op}+R_{\rm rel})$.
Then, if the optical pumping rate equals
the spin relaxation rate (between the ground sub-states),
the above-mentioned polarization $P_z=1/2$ can be achieved.

Actually, the above reduced Bloch equation is the Eq.\ (7) in the main text,
which is employed to solve the electron spin polarization vector
in the presence of magnetic field.
For the solving method and result, see description in the paragraph below Eq.\ (7).

\section{The issue of polarization cross talk of PBS}

The basic principle of the FR-based optical AM is letting
the probe light suffer an optical Faraday rotation (FR)
when passing through the vapor cell (in the AM),
then measuring the FR angle which allows to infer an unknown magnetic field.
Therefore, the ultimate limit of weak magnetic fields probe
is crucially affected by the performance of the PBS, as shown in Fig.\ 1(b).

Below we explain this important issue.
In real system, we can define and measure the ratio
of the different polarization light
(i.e. the $|H\ra$ and $|V\ra$ components)
exiting from the PBS, as shown in Fig.\ 5, as
\bea
V_{\rm m}
= \frac{\eta_R (I_H R_H + I_V R_V)}{\eta_T (I_H T_H + I_V T_V)}   \,.
\eea
Here $I_{H,V}$ are the intensities of the $|H\ra$ and $|V\ra$
components of the incident light (before entering the PBS);
$T_{H,V}$ and $R_{H,V}$ are the transmission and reflection coefficients;
and $\eta_{T,R}$ are the photo-electric converting coefficients
of the photo-detectors in the transmission and reflection branches.
In order to eliminate the effect of the nonideal
photo-electric converting coefficients, i.e., $\eta_T\neq\eta_R\neq 1$,
one can apply the strategy of {\it calibration} \cite{Van11,Wang13,Ben01,Win06,Yang19}.
That is, first, using a natural light as an incident light onto the PBS, one obtains
\bea
V_{\rm cal}
= \frac{\eta_R (R_H + R_V)}{\eta_T (T_H + T_V)}   \,.
\eea
Here the property that a natural light has
equal weight of $|H\ra$ and $|V\ra$ components has been considered.
Then, one can define the {\it calibrated ratio} as
$\widetilde{V}_{\rm m} = V_{\rm m} / V_{\rm cal}$ \cite{Van11,Wang13,Ben01,Win06,Yang19}.
Straightforwardly, we have
\bea
\widetilde{V}_{\rm m}
&=& \left[ \frac{\eta_R (I_H R_H + I_V R_V)}{\eta_T (I_H T_H + I_V T_V)}  \right]
\left[ \frac{\eta_R (R_H + R_V)}{\eta_T (T_H + T_V)}  \right]^{-1}   \nl
&=& \left( \frac{\delta_1 + V_0 R_V }{ T_H + V_0 \delta_2}  \right)
\left( \frac{\delta_1 + R_V}{T_H + \delta_2 }  \right)^{-1}   \,.
\eea
Here we introduced:
$V_0=I_V/I_H$, the ratio of the two polarization components;
and $R_H=\delta_1$ and $T_V=\delta_2$,
two (small) parameters to characterize the polarization cross talk.
In ideal case (theoretical model), there is no cross talk,
then the calibrated ratio $\widetilde{V}_{\rm m}$
recovers the true ratio $V_0$
in the incident light (before entering the PBS).
This actually suggests the basic principle
of determining the ratio $V_0$,
simply using the measured result of $\widetilde{V}_{\rm m}$.

\begin{figure}
  \centering
  \includegraphics[scale=0.5]{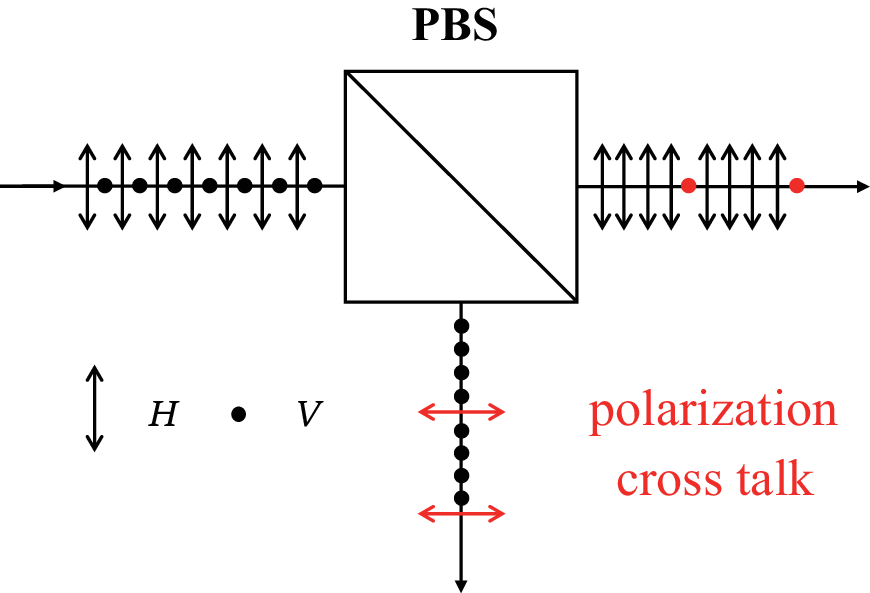}
  \caption{
Illustration of the polarization cross talk in a polarizing beam splitter (PBS).
In ideal case, the $|H\ra$ components
in the incident light fully transmit through the PBS,
while the $|V\ra$ components are fully deflected downwards.
In real case, there exists {\it polarization cross talk}:
a small portion of $|V\ra$ components transmit through the PBS,
and some $|H\ra$ components are deflected.     }
\label{fig2}
\end{figure}

\begin{figure}
  \centering
  \includegraphics[scale=0.45]{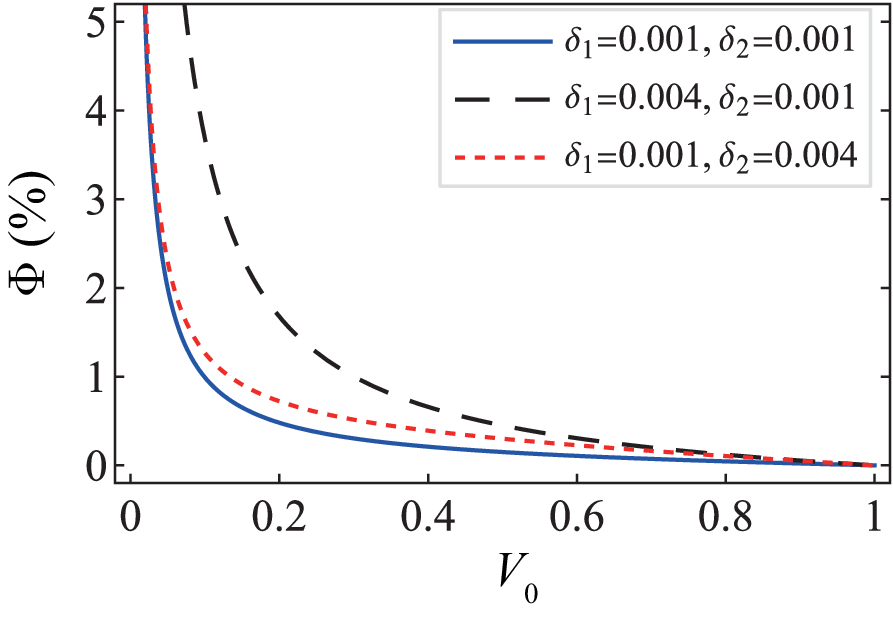}
  \caption{
Dependence of the $V_0$ estimation error ratio on the magnitude of $V_0$,
illustrating by a few cross-talk parameters $(\delta_1, \delta_2)$,
which are in the range of $10^{-3}\sim 10^{-2}$
in real systems \cite{Van11,Wang13,Ben01,Win06,Yang19,Bau00,Li23}.
$V_0=I_V/I_H=P_V/P_H$ is the ratio of the two polarization components
in the incident light.
It is clear that, with the decreasing of $V_0$,
the polarization cross talk will cause more serious problem,
even making the estimation completely failed.    }
\end{figure}

However, in real PBS system, inevitably,
there exists polarization cross talk, as schematically shown in Fig.\ 5.
This imperfection will affect the ultimate limit of $V_0$ measurement (estimation),
and has received extensive studies
in the context of polarization lidar \cite{Win06,Yang19,Bau00,Li23}.
To characterize the quality factor of the PBS,
we can introduce the error ratio
\bea
\Phi = \frac{|\widetilde{V}_{\rm m}-V_0|}{V_0}   \,.
\eea
In Fig.\ 6, we illustrate the dependence of this error ratio on $V_0$,
for a few polarization cross talk parameters,
which are in the range of $10^{-3}\sim 10^{-2}$
in real systems \cite{Van11,Wang13,Ben01,Win06,Yang19,Bau00,Li23}.
Importantly, it is clear that, with the decreasing of $V_0$,
the polarization cross talk will cause more serious problem,
even making the estimation completely failed.
In our case, for the FR-based full optical AM,
we just encounter the small $V_0$ problem,
since the weak magnetic field (to be probed)
only cause a very weak Faraday rotation (from $|H\ra$ to $|V\ra$),
which corresponds to a small $V_0$.
Therefore, the PSA strategy proposed in this work
is anticipated to be useful to
enhance the ultimate limit of weak magnetic fields probing,
since the postselected light
has a much larger $V_0$ (two or three orders of magnitude larger),
as shown in Fig.\ 2(b) and Fig.\ 3 in the main text.

\clearpage

\end{document}